# TOP-GAN: Label-free cancer cell classification using deep learning with a small training set

Moran Rubin[1,2], Omer Stein[2], Nir A. Turko[1], Yoav Nygate[1], Darina Roitshtain[1],
Lidor Karako[1], Itay Barnea[1], Raja Giryes[2], and Natan T. Shaked[1,*]

[1] *Department of Biomedical Engineering, Faculty of Engineering, Tel Aviv University, Tel Aviv 69978, Israel*
[2] *School of Electrical Engineering, Faculty of Engineering, Tel Aviv University, Tel Aviv 69978, Israel*
[*] *Corresponding Author e-mail: nshaked@tau.ac.il*

*Abstract*—We propose a new deep learning approach for medical imaging that copes with the problem of a small training set, the main bottleneck of deep learning, and apply it for classification of healthy and cancer cells acquired by quantitative phase imaging. The proposed method, called transferring of pre-trained generative adversarial network (TOP-GAN), is a hybridization between transfer learning and generative adversarial networks (GANs). Healthy cells and cancer cells of different metastatic potential have been imaged by low-coherence off-axis holography. After the acquisition, the optical path delay maps of the cells have been extracted and directly used as an input to the deep networks. In order to cope with the small number of classified images, we have used GANs to train a large number of unclassified images from another cell type (sperm cells). After this preliminary training, and after transforming the last layer of the network with new ones, we have designed an automatic classifier for the correct cell type (healthy/primary cancer/metastatic cancer) with 90-99% accuracy, although small training sets of down to several images have been used. These results are better in comparison to other classic methods that aim at coping with the same problem of a small training set. We believe that our approach makes the combination of holographic microscopy and deep learning networks more accessible to the medical field by enabling a rapid, automatic and accurate classification in stain-free imaging flow cytometry. Furthermore, our approach is expected to be applicable to many other medical image classification tasks, suffering from a small training set.

## I. INTRODUCTION

Cancer is a leading cause of death worldwide. Flow cytometry of body fluids obtained by routine medical tests can identify circulating tumor cells [1,2]. However, isolation of these cancer cells is laborious and typically yields uniformly round cells, which are hard to grade [3]. In flow cytometry for cell sorting, one evaluates cellular features through fluorescence markers [4]. However, fluorescent markers tend to photobleach, which damages the image contrast and the prognosis decisions [5]. The morphology and texture of cancer cells changes during cancer progression [6,7]. Without staining, however, biological cells are nearly transparent, resulting in a low image contrast.

An internal contrast mechanism that can be used when imaging cells without staining is their refractive index. The light beam passing through the imaged cells is delayed, since the cells have a slightly higher refractive index compared to their surroundings. Conventional intensity-based detectors are not fast enough to record this light delay directly. Phase imaging methods, on the other hand, use optical interference to record the delay of light passing through the sample, and thus they yield stain-free contrast in the image. Contrary to qualitative phase contrast methods, interferometric phase microscopy (IPM) yields the full sample wavefront, containing the optical thickness map or optical path delay (OPD) map of the cell, so that on each spatial point of this map, OPD is equal to the integral of the refractive index values across the cell thickness [8]. In addition to contrast obtained on all cell points without staining, IPM allows calculating quantitative parameters, such as cell volume and dry mass, which were not available to clinicians so far [9]. In the last years, we made significant efforts to make these wavefront sensors affordable for clinical use [10-13] by attaching a portable interferometric module to exit port of an existing in clinical microscope, making this technology accessible and affordable to the clinicians' direct use.

In contrast to previous works presenting statistical discrimination of holographic data of cells [14-17], the advantage of machine learning classifiers is that they can work on multi-feature space, or even on unknown feature space, for classification between the groups examined.

Combining holographic microscopy with recent developments in the field of machine learning enables automatic label-free analysis of large amounts of cells which assist with the classification of different types of cells [18-28]. Furthermore, various combinations of machine learning and holographic microscopy have been proposed lately (e.g. [29-32]).

The fact that IPM can now be implemented in clinical settings and provide the clinician tens of new parameters extracted from the cell OPD map gives rise to the question of how to account for all parameters together in order to classify the cells. Simple machine learning approaches can weigh the various parameters, for example by principle component analysis (PCA), followed by a support vector machine (SVM) classifier [21].

Instead of manually designing parameters to be extracted from the OPD maps, an alternative approach is to apply machine learning techniques directly on these maps with the goal of cell classification. This approach is more global, since it creates hidden parameterization that might be missed by the manually designed parameters, and thus, although more computationally heavy, it is expected to yield better classification results.

In the past years, the concept of deep convolutional neural networks (CNNs) has revolutionized the field of image recognition and classification [28,33,34]. When the number of parameters in the network is large, more samples are required for training in order to avoid overfitting situations [35]. One major challenge when building CNNs is that it is frequently hard to acquire very large number of classified training examples. This problem is common to other medical classification tasks solved with deep learning, and researchers thrive to suggest new approaches that allow applying deep learning with limited number of examples [36-38].

A simple possible solution to the problem of a small training set is to expand the training set by using data augmentation [32,39]. However, the new information gained from this process might be minor and completely ineffective for many types of data. An alternative solution to the problem of a small training set is transfer learning. It can be used in the case of well-known deep learning networks that were previously trained on large general training sets [40-42]. By changing the last layers of these pre-trained networks, researchers have shown the possibility of receiving high accuracy results after training with a small data set, specifically for medical images [41]. Another approach is based on generative adversarial networks (GANs). This is an unsupervised learning strategy for generating new data points for a given large unclassified dataset [43]. This model is composed of two individual networks that compete with each other and by this learn to synthetically create additional reliable images, which increase the available training set [44,45]. Alternatively, GANs can be used as semi-supervised learning techniques that utilize a large number of unclassified images in order to achieve better classification results with a small number of classified images of the same type [46,47]. These approaches, however, cannot help if neither a sufficient number of classified nor unclassified training images from the same type are available. Another approach is to train GAN model on a large set of unclassified images, and then fine-tune one of its networks on a smaller set of classified images [48,49]. To the best of our knowledge, this approach has not been adapted so far for medical imaging.

In the current paper, we perform classification of quantitative phase images of healthy and cancer cells and of primary cancer and metastatic cancer cells. For these tasks, we successfully apply various machine learning methods, one of which copes with the problem of a small training set by using another set of unclassified images, and is used here in the first time for medical imaging. This method, called transferring of pre-trained generative adversarial network (TOP-GAN), exploits the large amount of unclassified data from another set of examples in order to compensate for the lack in classified data. In order to do so, we combine between transfer learning and GAN. TOP-GAN uses GAN in order to train a discriminator network on a large number of unclassified images from one type. Then, by changing the last layer of the discriminator with new un-trained layers, we transform the discriminator network into a new classifier network that can be retrained on a small number of classified images from another type. The transformation of the network is done using transfer learning; hence, we train the GAN model on one type of data set, e.g. unclassified sperm cell OPD images, and then implement this information for classification of a smaller classified data set of another type of data set, e.g. classified cancer and non-cancer cell OPD images. To the best of our knowledge, this is the first time that the discriminator network is being used as a transfer learning network for medical imaging in general and for stain-free cell classification in particular.

In this paper, we examine the classic CNN abilities on classification two datasets of the cell OPD profiles and the abilities of different machine learning methods, including TOP-GAN, to improve the classification accuracy of the classic CNN in the case of a small and complex dataset.

By comparing our hybridized approach to other common approaches for medical image classification, we show that TOP-GAN yields a more efficient classifier, which can achieve better classification results with a smaller classified training set. Furthermore, this method shown to be much more robust to the selection of the training set than other deep learning approaches.

## II. MATERIALS AND METHODS

We imaged four cell lines. Each pair of cells is taken from the same individual and the same organ, and the cells are imaged without staining while being unattached and thus mostly round, as is the case in imaging flow cytometry, which makes it hard to classify. The first set represents a less complex task, discriminating between healthy skin and cancer cells, while the second set represents a more complex task, classification between primary cancer and metastatic cancer cells.

### A. Preparation of Human Cancer Cells

We performed measurements for two pairs of isogenic cell lines: 1) Hs 895.Sk (healthy skin) and Hs 895.T (melanoma), 2) SW 480 (colorectal adenocarcinoma colon cells) and SW 620 (metastatic from lymph node of colorectal adenocarcinoma cells). The cells were purchased from the ATCC, and each of the cell line comparisons originated from the same individual. The first pair of the cells was used for the classification between healthy skin cells and cancer cells, and the second pair of cells was used for classification between primary cancer cells and metastatic cancer cells.

The complete growth medium used for the Hs cells was Dulbecco's Modified Eagle's Medium (DMEM), supplemented with 10% Fetal Bovine Serum (FBS) (BI) and 2 mM L-glutamine (all purchased from biological industries, Beit Haemek, Israel). The complete growth medium used for the SW cells was BI Roswell Park Memorial Institute (RPMI) 1640 Medium without L-glutamine supplemented with 10% FBS and 2 mM L-glutamine (all purchased from biological industries, Beit Haemek, Israel).

All cell lines were incubated under standard cell culture conditions at 37°C and 5% $CO_2$ in a humidified incubator until 80% confluence was achieved. Prior to the imaging experiment, the cells were trypsinized for suspension, supplemented adhesive chamber, volume 18 μL, 13 mm diameter × 0.15 mm thickness, 1.5 mm ports diameter, Sigma Aldrich SN. GBL611101) attached to a cover slip. This chamber induced a contrast thickness value on the entire imaged sample, which is important for the flatness of the final phase map. Then, all cell lines were imaged by IPM.

*B. Preparation of Human Sperm Cells*

We obtained sperm cells from six human donors at their 20s. The study was approved by the institutional ethics committee of Tel Aviv University. All sperm donors signed a written informed consent form. After ejaculation, the semen was liquefied at room temperature for 30 minutes and then the sperm cells were isolated using the PureCeption Bi-layer kit (Origio, Målov, Denmark) in accordance with manufacturer's instructions. The semen and non- spermatozoa cells were discarded, and the pellet that include sperm cells was resuspended in 5 mL of modified human tubal fluid (HTF) medium (Irvine Scientific, California) and centrifuged at 500 g for 5 minutes. For fixation the HTF medium was discarded, the pellet was resuspended with 0.1 ml HTF medium and then 10 ml fixative solution (3:1 methanol to acetic acid) was added drop by drop. After 5 minutes at room temperature, the cells were centrifuged at 800 g for 5 minutes, the supernatant was discarded, and the pellet was resuspended in 0.2 mL of fixative solution. 10 μl of the fixed cells were smeared on a 60 × 20 mm #1 cover slips and put to dry overnight to ensure the evaporation of the fixative solution. Cells were then imaged by IPM.

*C. Interferometric Phase Microscopy (IPM)*

In order to obtain the OPD topographic maps of the cells, we acquired off-axis image holograms. For imaging Hs cells, we used the system mentioned in [21]. For the SW cells, we used flipping interferometry module [12], connected to an inverted microscope. Flipping interferometry is a compact and portable module, so it can be attached into existing clinical microscopes, signifying its advantage for clinical imaging flow cytometry. The microscope was illuminated by a coherent laser (Helium-Neon, 632.8 nm), and a microscope objective (Mitutoyo, 50X, 0.55 NA) was used for imaging. The off-axis image holograms were created on the digital camera (Thorlabs, DCC1545M) positioned right after the interferometric module. In this module, a beam splitter splits the beam into a reference beam, which is flipped by a retro-reflector, and a sample beam, which is back-reflected by a mirror. This configuration requires half of the optical field of view to be empty from sample details. This flipping interferometric geometry is specifically useful for microfluidic channels, since it is easier to make sure that the sample beam half is positioned on the area of flowing cells and the other half of the beam, dedicated for the reference, is positioned on the bare glass of the channel. Therefore, the flipping interferometry module can deal with non-sparse samples without creating ghost images, in contrast to modules based on shearing interferometry (e.g., [50]). The OPD maps were extracted digitally from the acquired off-axis holograms as explained at the beginning of the Quantitative Phase Reconstruction section.

*D. Quantitative Phase Reconstruction*

All image analysis procedures were carried out using Matlab (R2017b). We extracted the OPD maps from the off-axis holograms of the cells by using digital spatial filtering [51]. The algorithm included a 2D Fourier transform, filtering one of the cross-correlation terms containing the complex wavefront of the sample, and an inverse 2D Fourier transform. The resulting matrix argument was the wrapped phase of the sample. We subtracted from the wrapped phase map of the sample a phase map, which was extracted from a reference hologram (without the sample present), in order to overcome stationary aberrations and field curvatures. We have then created the unwrapped phase map, by using the unweighted least squares phase unwrapping algorithm, and multiplied it by the wavelength of the source divided by $2\pi$ in order to create the quantitative OPD map of the sample, which is defined as follows:

$$OPD_c(x,y) = [\overline{n_c}(x,y) - n_m] \times h_c(x,y), \qquad (2)$$

where $n_m$ is the refractive index of the medium, $h_c$ is the thickness profile of the cell, and $\overline{n_c}$ is the cell integral refractive index, which is defined as follows:

$$\overline{n_c}(x,y) = \frac{1}{h_c}\int_0^h n_c(x,y,z)dz. \qquad (3)$$

The reconstructed OPD maps, containing multiple cells, were segmented by thresholding, in order to separate the cells from the background, and then cropped into images of single cells. The isolated cells were computationally aligned at the center of square backgrounds for further analysis. Using the method described above, we created a data set of RGB images with size 128x128x3 [52] that described the OPD information of the individual cell area only. This data set was the input to the deep learning networks.

*E. K-nearest neighbor (KNN)*

KNN is a naïve algorithm that does not make any assumptions on the data distribution. In the case of image classification, the algorithm creates an n-dimensional space that is based only on the features extracted. In our case, however, we have chosen to implement KNN in the image space by calculating the Euclidean difference (L1) between the test image and any training images in order to find the K nearest neighbors. This can define the complexity of each class and the similarity between different images in the same class. This classifier provides good classification when the number of classified training images is high relative to the class complexity [53]. In our case, K = 9 yielded the best accuracy. There is a direct relation between the performance of a CNN and the performance of a simple KNN [54]. For the implementation of the KNN, we use the TensorFlow framework [55].

*F. Convolutional Neural Network (CNN)*

We use the convolutional CNN illustrated in Fig. 1. The architecture of this network is composed of four convolution layers followed by three fully connected layers with sizes of 100, 100, and 2, respectively. In order to reduce the chances of overfitting, each one of the first two fully-connected layers is followed by a dropout layer with a probability of 0.5 during training [56]. The input of the network is an RGB image with a size of 128x128x3 pixels, which describes an OPD map of a cell, and the output is a binary decision value that signifies if the image contains a healthy skin or a cancer cell for the first experiment, and a primary cancer or a metastatic cancer cell for the second experiment. The architecture of the conventional CNN is based on the discriminator network from the DCGAN model, thus a sigmoid function is used in the last layer. For the implementation of the CNN architecture, we use the TensorFlow framework [55]. For training, we use an Adam optimizer with a learning rate of 0.00001 and adaptive momentum with the parameters: $\beta_1 = 0.6, \beta_2 = 0.99$. The network is trained for 900 epochs or until convergence, whichever comes first.

*G. MOBILE-NET for Transfer Learning*

MOBILE-NET is trained on the ImageNet of natural images [42]. It is lightweight in its architecture, and therefore is useful for transfer learning with a small training set. We then use this network of non-medical images for medical image analysis by fine-tuning the previously trained network with classified OPD images of biological cells, we are using the information and general features collected from previous training in order to improve the relevant classification task. We implement the MOBILE-NET model using the TensorFlow framework [55].

Note that we have also implemented transfer learning by using a larger network, VGG16, which has also been previously trained on ImageNet database, and have compared it to the other methods. However, it has yielded worse results than MOBILE-NET presented above

*H. Data Augmentation*

The standard solution to reduce overfitting is data augmentation that enlarges size of the data sets. However, the success of data augmentation process is mainly dependent on the data itself. As described in [39], classic data augmentation techniques include rotation, flipping and scaling of the input images. In order to avoid changes in the characteristics and morphology of the cells, we perform OPD image augmentation by combining between flipping and rotation (without scaling). The transformation of the OPD maps is done by using Matlab (R2017b). By doing so, we manage to increase our data set by eight times.

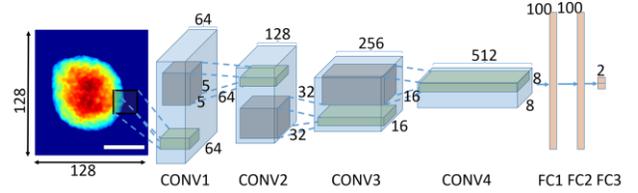

**Fig. 1.** The architecture of the CNN and of the TOP-GAN for classification of OPD images of cancer cells.

*I. Generative Adversarial Network (GAN)*

GAN is an unsupervised learning strategy that utilizes a given large unclassified samples to generate new data points (e.g. images) based on the same distribution [43]. This model is based on training two individual convolutional neural networks, generator and discriminator. The generator network attempts to synthetically create a realistic images with a similar distribution to the real images, in order to 'fool' the discriminator network, whereas the discriminator have to identify correctly which input image is real and which one is fake. In our research, we have based our model on the architecture of deep convolutional GAN (DCGAN), as illustrated in Fig. 2(a) and on the suggested guidelines for stable GAN training proposed in [44]. The architecture of both networks of the GAN model, generator and discriminator, is illustrated in Fig. 2(b) and Fig. 2(c), respectively.

The generator network takes as an input a vector with a size of 100 elements that are sampled from a random normal distribution. After propagating through the network, it outputs an RGB image that describes an OPD map of a sperm cell with a size of 128x128x3 pixels. The network architecture consists of a fully connected (FC) layer reshaped to a size of 8x8x512 and four deconvolution layers with a stride of 2 and a kernel size of 5x5. Batch-normalization layers and a ReLU activation functions are applied to all layers except the output layer which uses a tanh activation function [44].

The discriminator network has a classic CNN architecture. The input of the network is an RGB image with a size of 128x128x3 pixels that describes an OPD map of a cell. The output of the network is a binary decision value that signifies if the image is a real OPD map or a fake one. The network architecture consists of four convolution layers with a stride of 2 and a kernel size 5x5, and a fully connected layer. A batch normalization layer is applied to each convolution layer except the input and the output layers. Leaky ReLU activation functions, with a slope set to 0.1, are applied to all convolution layers except the output layer, which uses the sigmoid function [44]. The Sigmoid function gives the probability (0,1) of the image to be 'fake' or 'real' (respectively).

The adversarial networks are trained by optimizing the follow loss functions:

Discriminator:

$$max_D E_{x \sim P_{data}}[log(D(x))] + E_{x \sim P_z}[log(1 - D(G(z)))], \quad (4)$$

Generator

$$max_G E_{x \sim P_z}\left[log\left(D(G(z))\right)\right], \quad (5)$$

where $D(x)$ and $D(G(z))$ describe the output of the discriminator in case the input is a real image or a generated one respectively, and where $P_{data}$ and $P_z$ describe the distribution of the input image in the case of real image or a generated one, respectively. The discriminator is trained to maximize $D(x)$ for images with a distribution of $x$ that is similar to the distribution of the input samples, $P_{data}$. Furthermore, it is also trained to minimize $D(x)$ for images with a different distribution. The aim of the generator is to create generated images $G(z)$ with a distribution that is similar as much as possible to the distribution of the input samples in order to 'fool' the discriminator during training. Therefore, based on [43], the generator is trained to maximize $D(G(z))$.

In order to train both the generator and the discriminator networks, we use 2762 different OPD images of human sperm cells captured in our lab by IPM. In order to increase the data set size, we use a data augmentation method, which increases the data set by eight, as will be describe next. During the training process, we use mini-batches of 64 OPD images of human sperm cells, and 64 samples of random noise, Weights are initialized to a zero-centered normal distribution with standard deviation of 0.02. We apply an Adam optimizer [xx] with adaptive momentum with the parameters: $\beta_1 = 0.5, \beta_2 = 0.99$ and a learning rate of 0.0002 for 75 epochs. We implement the GAN model using the TensorFlow framework [55].

*J. TOP-GAN*

The TOP-GAN is implemented by performing transfer learning with the previously trained discriminator from the GAN model described in the previous section. After the GAN is trained on thousands of unlabeled OPD images of human sperm cells, both the generator and discriminator networks have already learned most of the important features that characterize OPD images of biological cells. We thus use the discriminator network from the first section, which has been already trained on sperm cells, and switched its last layer with three un-trained fully connected layers. In fact, we have created a classifier that is based on the architecture and knowledge of the discriminator described from the previous section. The classifier architecture is the same architecture as the convolutional CNN, illustrated in Fig. 1, and describes four pre-trained convolution layers followed by three un-trained fully connected layers with sizes of 100, 100, and 2, respectively. Also here, each one of the first two un-trained fully-connected layers added to the network was followed by a dropout layer with a probability of 0.5 during training [56]. Like the original discriminator network, the input of the TOP-GAN is an RGB image with a size of 128x128x3 pixels, which describes an OPD map of a cell. However, the output is a binary decision value that signifies if the image is an image of a healthy skin and cancer cell for the first experiment, and preliminary cancer and metastatic cancer cell for the second experiment. Similarly to the discriminator, also the TOP-GAN uses a sigmoid function in the last layer. For the implementation of the TOP-GAN architecture, we use the TensorFlow framework [55]. For training, we use an Adam optimizer with a learning rate of 0.00001 and adaptive momentum with the parameters: $\beta_1 = 0.6, \beta_2 = 0.99$. The network is trained for 900 epochs or until convergence, whichever comes first.

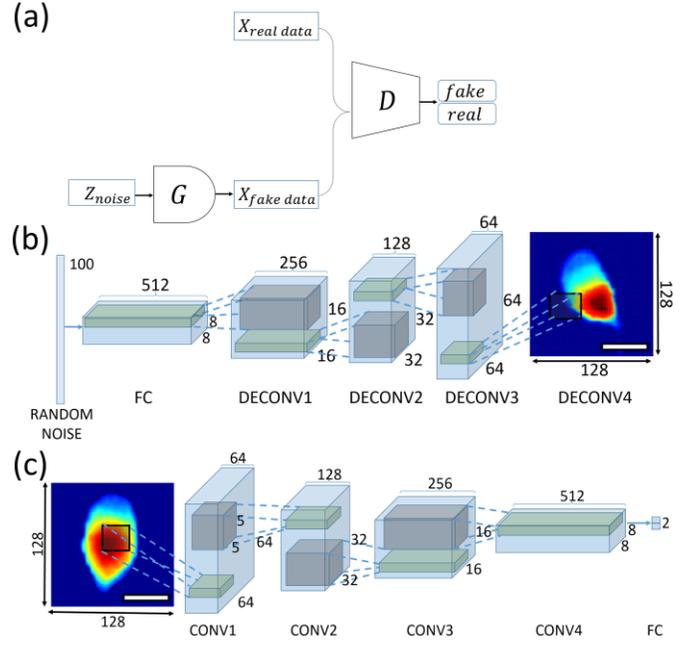

**Fig. 2.** (a) DCGAN architecture. (b) Generator network architecture when generating OPD images of sperm cells. (c) Discriminator network architecture when training on real and generated OPD images of sperm cells.

### III. RESULTS

We used IPM to acquire off-axis digital holograms of individual unstained cells for the reconstruction of their OPD maps. To this end, we used an external module that can be implemented in a clinical setting (see Materials and Methods). Figure 3 presents the OPD maps of unattached cells of type: Hs 895.Sk (skin), Hs 895.T (melanoma), SW 480 (colorectal adenocarcinoma colon) and SW 620 (metastatic colorectal adenocarcinoma colon), respectively. These figures demonstrate the difficulty of visually finding significant differences between these groups of cells with a naked eye, even with quantitative phase imaging. This is due to the fact that typically isolated, unattached cells are mostly round-shaped, and thus are fairly similar to one another [3]. This is the case in imaging flow cytometry.

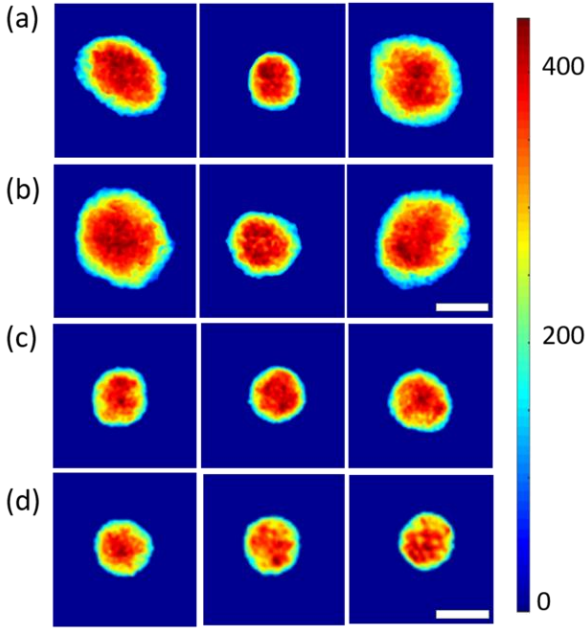

**Fig. 3.** Quantitative optical path delay (OPD) maps of unattached cancer cells, demonstrating the similarity between these groups. (a) Hs 895.Sk (healthy skin cells), (b) Hs 895.T (melanoma cells), (c) SW 480 (colorectal adenocarcinoma colon cells), (d) SW 620 (metastatic colorectal adenocarcinoma colon cells). Color bars represent OPD values in nm.

Our goal is to build a robust, automatic network that can accurately classify between two stages of the cancer cells, even when the number of available classified samples is low. In order to do so, we examine two different tasks with different levels of complexity. In the first experiment, we classify melanoma cells (Hs 895.T) and healthy skin cells (Hs 895.Sk), and in the second experiment we classify metastatic cells (SW 620) and primary cells (SW 480) of colorectal adenocarcinoma colon imaged during flow. For both experiments we first use the classic CNN for classification, and then try to improve its results by different machine learning classification methods, including TOP-GAN.

To evaluate the classification performance of each method, we use a total classification accuracy measure. Additionally, we calculated the sensitivity, and specificity measures for each experiment. All measures are presented in the following equations:

$$\text{Total Accuracy} = \frac{TP+TN}{TP+TN+FP+FN}, \qquad (5)$$

$$\text{Sensitivity} = \frac{TP}{TP+FN}, \qquad (6)$$

$$\text{Specificity} = \frac{TN}{TN+FP}, \qquad (7)$$

where true (T) and false (F) represent images classified correctly or misclassified, respectively, while positive (P) and negative (N) represent the different categories (e.g. healthy/cancer, primary/metastatic). In order to maximize the sizes of the training and testing sets, and in order to test the effect of the chosen training images on the success of the experiment, we perform five-fold cross-validation. This process includes five iterations, where during each of them we choose different images for training and images for testing. Each training procedure is performed separately, and does not influence the results of the other four training steps. The final accuracy, specificity and sensitivity obtained from this training process are the mean value from all five iterations for each task. The comparison between the different methods (except for the KNN) is illustrated in Fig. 4(a) and Fig. 4(b), for the first and second experiments, respectively. We also checked the robustness of each method to the selected training set by inspecting the range of resulting accuracies, as shown in Fig. 4(c) and Fig. 4(d). Tables 1 and 2 demonstrate the best performance of the TOP-GAN in terms of sensitivity, specificity and area under curve (AUC), for the smallest training set as possible and compare it to other methods (CNN, CNN-WITH-AUG, and MOBILE-NET). Table 1 summarizes the mean results from 5 different folds for specificity, sensitivity and AUC for the first experiment (healthy vs. cancer) by using 10 images for training and 192 images for testing. Table 2 summarizes the same for the second experiment (primary cancer vs. metastatic) by using 236 images for training and 40 images for testing. All results for all methods are calculated on the same training and testing sets. All training processes have been performed using the Google Cloud Platform on the NVIDIA's Tesla P100 GPU.

All acquired and synthetic OPD images used for training have the same noise characteristic of OPD images acquired under coherent illumination, rather than incoherent illumination, presenting the worst case scenario for spatial noise. We therefore did not choose to repeat the experiment for cases of higher synthetic noise. The following classification methods have been applied:

*A. K-nearest neighbor (KNN)*

We first use a naïve KNN classifier in order to predict the complexity of each experiment. The following results are obtained:

***Healthy vs. cancer cells:*** By using a training set of 162 images of melanoma cells and healthy skin cells, the KNN algorithm is able to reach 88% accuracy. This high accuracy describes the low complexity of this dataset, where the images associated with the same class are very similar to each other [53]. However, when decreasing the training set, the value of the mean accuracy decreased as well. For example, by training with 10 images, the KNN algorithm reaches only 69.36% mean accuracy.

***Primary vs. metastatic cancer cells:*** By using a training set of 236 images of primary and metastatic adenocarcinoma cells, the KNN algorithm is able to reach only 82.02% accuracy, lower than the results in the experiment of the melanoma cells even when the training set now is larger. We thus conclude that images that are associated with the same class are less similar each other than in the first experiment, making it harder to classify them based on a small training set. When decreasing size of the training set, the value of the mean accuracy decreases even further, reaching a mean value of less than 70% with 60 images for training. Furthermore, the KNN results in this experiment have been very sensitive to the training set selection. This fact proves the major influence of the images selection to the performance of the KNN algorithm.

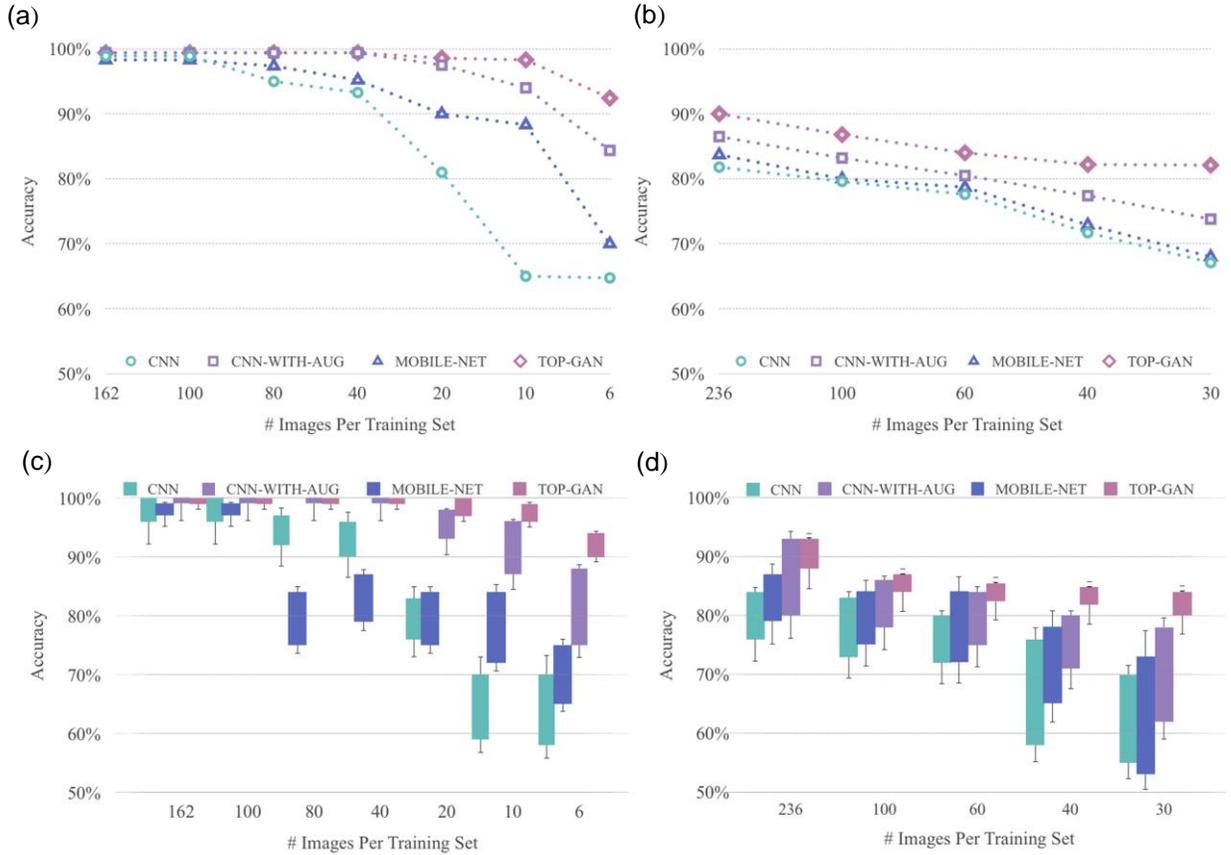

**Fig. 4.** The mean accuracy result calculated on 5 different training sets for different training sizes and different methods (CNN, CNN-WITH-AUG, MOBILE-NET and TOP-GAN) for classification of melanoma cells and healthy skin cells (a) and for classification of primary and metastatic adenocarcinoma cells (b). The range of the accuracy results (percentage) for different sizes of training set. Each color in the graph represents a different method (CNN, CNN-WITH-AUG, MOBILE-NET and TOP-GAN) for classification of melanoma cells and healthy skin cells (c), and for classification of primary and metastatic adenocarcinoma cells (d).

**Table 1**
Performance comparison between the different methods (CNN, CNN-WITH-AUG, MOBILE-NET, TOP-GAN, and TOP-GAN-AUG) for Hs 895.Sk (healthy) vs. Hs 895.T (cancer) with 10 images per train and 192 images per test

| Method | Sensitivity (%) | Specificity (%) | AUC |
|---|---|---|---|
| TOP-GAN-AUG | **98.96** | **99.61** | **0.995** |
| TOP-GAN | 98.08 | 96.76 | 0.991 |
| MOBILE-NET | 89.21 | 93.6 | 0.949 |
| CNN-WITH-AUG | 88.28 | 93.2 | 0.952 |
| CNN | 65.98 | 66.44 | 0.717 |

**Table 2**
Performance comparison between the different methods (CNN, CNN-WITH-AUG, MOBILE-NET, TOP-GAN, and TOP-GAN-AUG) for SW 620 (primary cancer) vs. SW 480 (metastatic) with 236 images per training and 40 images per test

| Method | Sensitivity (%) | Specificity (%) | AUC |
|---|---|---|---|
| TOP-GAN-AUG | **93.01** | **93.31** | **0.947** |
| TOP-GAN | 90.61 | 92.94 | 0.942 |
| MOBILE-NET | 85.01 | 79.21 | 0.881 |
| CNN-WITH-AUG | 86.01 | 83.21 | 0.892 |
| CNN | 82.32 | 78.81 | 0.878 |

*B. CNN*

The following results have been obtained by applying a conventional CNN with the same architecture as the TOP-GAN (illustrated in Fig. 1).

*Healthy vs. cancer cells:* Our CNN is trained as a binary classifier on a training set of 162 classified OPD images of individual melanoma cells and healthy skin cells. The performance of the trained CNN is tested on 40 new OPD maps (20 images of melanoma cells and 20 images of healthy skin cells). As usual, the test images have never been seen previously by the network. The diagnostic accuracy is surprisingly high, resulting in a mean accuracy of 98.9%. The accuracy calculated is defined in (5), where Hs 895.T is defined as 'positive' and Hs 895.Sk as 'negative'. The high accuracy results for a relatively small training set, with only 162 classified images, can be explained by Ref. [54] that demonstrates the connection between the high accuracy results of the naive KNN algorithm and the results of a deep neural network. While reducing the size of the training set, the performance of the conventional CNN drops as well. For example, by training the CNN with 10 images and testing it on 192 images, the network barely reaches to 65% accuracy.

*__Primary vs. metastatic cancer cells:__* Our CNN is trained as a binary classifier on a training set of 236 classified OPD images of both individual primary adenocarcinoma and metastatic cells. The performance of the trained CNN is tested on 40 new OPD maps of both primary (20 images) and metastatic (20 images). The test images have never been seen previously by the network. The accuracy calculated is defined in (5), where SW 620 is defined as 'positive' and SW 480 as 'negative'. Also in this experiment, we have created five different training sets in order to evaluate the influence of the training images on the performance of the network, and the accuracy presented is the mean accuracy from all five iterations. In this experiment, the mean accuracy reaches 81.8%, much lower than the first experiment, even when using a larger number of classified images for training. Again, by looking at the result of the KNN algorithm for the same number of cells, we find a positive correlation between the success of the KNN and the success of the CNN, which coincides with previous results [54]. Like in the first experiment, as the size of the training set decreases, the accuracy of the classic CNN decreases as well, reaching 72% accuracy with 40 images for training. Furthermore, the selection of training set has significant effect on the results, as can be seen in Fig. 4(d).

*C. MOBILE-NET*

In these experiments, we take a MOBILE-NET [42] that has been trained on non-medical data, and fine-tune it on our classified training sets that include OPD images of biological cancer cells from different stages.

*__Healthy vs. cancer cells:__* The performance of MOBILE-NET was examined on the same data sets as for the conventional CNN, while reaching higher accuracy for smaller training size. As shown in Fig. 4(a), MOBILE-NET was able to maintained accuracy of 90% for only 20 images for training and was able to increase the accuracy performance of the conventional CNN by 23% for training of 10 classified images, reaching for 88% accuracy. However, this method was sensitive to the selection of training images, as can be seen in Fig. 4(c).

*__Primary vs. metastatic cancer cells:__* As shown in Fig. 4(b), MOBILE-NET was less effective in the second experiment (primary vs. metastatic), and was able to increase the accuracy of the conventional CNN with only 1% of mean accuracy for the same training set of 236 classified images. As the training set size decreases, the performance of the MOBILE-NET was decreased as well, and reached similar results as the conventional CNN. Also here, we can see that this method was highly influenced from the selected training set, as illustrated in Fig. 4(d). The resulting accuracy changed in the range of almost 20%, in the case of 30 training images, based only on the selection of the training set.

*D. CNN-WITH-AUG*

By combining between rotation and flipping, our training sets artificially increased by eight times. Following this process, the accuracy of the classification process has increased as well, even while using a smaller training set, as shown in Fig. 4(a) and Fig. 4(b).

*__Healthy vs. cancer cells:__* In the first experiment, while training on only 10 images and testing the result on the other 192 available images, the augmentation process increased the accuracy from 65% to 94%. Indeed, for a less complex task, data augmentation can improve the results.

*__Primary vs. metastatic cancer cells:__* In the second experiment, which is more complex (as can be concluded from the KNN results), the improvement in the results when using the data augmentation method is lower, improving from 81.8% to 86.5%. As shown in Fig. 4(d), this method was also very sensitive to the selection of training image, with accuracy range of 80%-93% for the same training set size.

*E. TOP-GAN*

TOP-GAN is meant to cope with a situation where only limited data sets of classified (known-class) images are available (classified OPD maps of cancer cells in our case) by creating a transfer learning network that fits another types of unclassified (unknown-class) images (unclassified OPD maps of sperm cells in our case). Our classification model is constructed by first building a GAN, based on the architecture of the deep convolutional GAN (DCGAN) model described in [44] and illustrated in Fig. 2(a). In our case, since thousands of unclassified OPD of sperm cells are available to us, and only a few tens or hundreds of classified OPD maps of cancer cells, the GAN training is performed by using a large number of OPD maps of unclassified sperm cells, and not the cancer cells.

The discriminator network that is pre-trained on OPD images of sperm cells becomes a transfer learning network for the classification of other biological cells, such as cancer cells. This technique is based on the idea that most OPD images of biological cells, especially in flow cytometry, appear highly morphologically similar. Therefore, the general features that characterize OPD maps of cancer cells are similar to those that characterize OPD maps of other biological cells, like sperm cells. Thus, by using GANs to pre-train the discriminator network on a large number of unclassified OPD maps of sperm cells, we allow the network to learn most of the generic features that characterize OPD maps of biological cells before using the small classified dataset. Note that in standard transfer learning, where the weights of a pre-trained network on another class with many examples are used, there is a need to have the data of the other class also classified. In TOP-GAN, on the other hand, we just need unclassified examples from the other class.

During the GAN training, the generator learns the features that characterize the OPD maps of sperm cells and is able to generate new OPD images with the same data distribution as the real OPD images. In Fig. 5(a), the images generated by the generator are illustrated for different stages of the GAN training process (from left to right). As the training process advances, the generated images become more and more similar to the real OPD images human sperm cells (see Fig. 5(b)). It is important to note that the GAN model is based on an adversarial process. Therefore, we can say that while the generator abilities in creating new images improve as the training progresses, the abilities of the discriminator network in classifying between these generated images and the real images improves as well. In fact, at the end of the training process, we can say that both networks, the generator and the discriminator, become familiar with the different features that characterize the OPD images of the sperm cells. After this preliminary training of the discriminator, we perform transfer learning, replacing the last layer of the discriminator with three new untrained fully connected layers, and change the classification labels from 'fake' and 'real' images to 'positive' and 'negative' images, according to our classification task.

Thus, by using a pre-trained network, we manage to create a new classifier that is more efficient in classifying a small amount of unseen biological cells, because most of its layers have already been trained on other data sets.

*Healthy vs. cancer cells*: In this case of low complexity data set, there is no need to use TOP-GAN (conventional CNN reached accuracy of 98.9%). However, if a smaller training set is available TOP-GAN should be used. For example, by training on just 10 classified images (5 melanoma cells and 5 healthy skin cells) TOP-GAN is able to still maintain a mean accuracy of more than 98%, higher than any of the classification method presented before. Additionally, in Table 1, we can see that also for sensitivity, specificity and area under curve (AUC), TOP-GAN yields the best results from all the classification methods.

*Primary vs. metastatic cancer cells:* In the second experiment, where cells were imaged during flow, the accuracy results of the conventional CNN were low. In this case, TOP-GAN reaches accuracy of 89.12% by training on the same 236 images as the CNN, presenting a significant improvement. Even after reducing the training sets massively, down to only 60 images for training, the TOP-GAN is still able to yield more than 83% accuracy. Additionally, in Table 2, we see that TOP-GAN yields the best sensitivity, specificity and AUC results. Furthermore, it was noticeable that while other methods yield a wide range of possible accuracy values for the same training size, based only on the selection of the training set, as can be seen in Fig. 4 (d), TOP-GAN yields a much more stable accuracy value for each training size. This signifies another advantage of TOP-GAN, namely being more robust to the selection of the training set.

Note that though our model is over-parameterized and reaches almost zero error on the training set, we believe that it does not suffer from overfitting the small data used due to the following reasons: (i) All state-of-the-art networks achieve almost zero error on the same train data. Yet, this memorization does not contradict their ability to generalize well; (ii) All the methods we compare to attain zero training error. Still, our performance on the test set is better; (iii) We use the same architecture for both the Hs 895.T vs. Hs 895.Sk databases, and the SW 620 vs. SW 480 datasets. For the second, the test error is far from 100%, so it is clear that our model is not overfitting of the test data. We thus conclude that the very high accuracy in the first case is just due to the structure of the data. Indeed, the KNN performance reported in the paper for each of the two cases explains the gap between the two experiments [53,54].

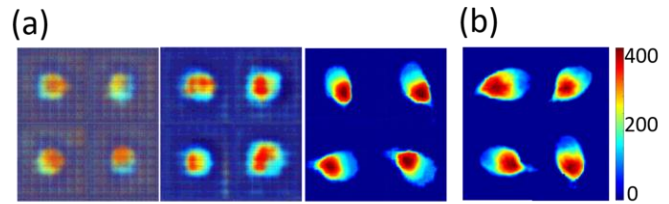

**Fig. 5.** The generated images from the generator output during training on OPD images of sperm cells in comparison to the real OPD images of the human sperm cells. (a) The generated images from left to right after 10, 20, and 75 epochs, respectively. (b) The experimentally acquired OPD images of sperm cells. Color bars represent OPD values in nm.

*F. TOP-GAN-AUG*

Finally, by combining data augmentation with TOP-GAN method, we are able to improve the results even further.

*Healthy vs. cancer cells*: For the first experiment, by training only 10 classified images and test them on all other 192 images, we reach 99.054% accuracy.

Furthermore, for this best performance method, Table 1 presents sensitivity of 98.96%, specificity of 99.61%, and AUC of 99.51%.

*Primary vs. metastatic cancer cells:* For the second experiment, by training with 236 classified images and test them on 40 images, we reach 90.6% accuracy.

Furthermore, for this best performance method, Table 2 presents sensitivity of 93.01%, specificity of 93.31%, and AUC of 94.73% are obtained.

## IV. DISCUSSION

We have suggested a method for an automatic classification of label-free individual cells based on the combination of deep learning and holographic microscopy that can be used even with a small classified training set, given that a large unclassified set of other biological cells is available. In our case, a large amount of data from one type (sperm cells) and small amount of data from other type (cancer cells) are available. Note that for this case, we have provided a comparison between many machine learning approaches. However, we have not provided comparison to methods that require a large amount of unlabeled data, and a small amount of labeled data from the same type, since this was not the case in our paper. We demonstrate that by using a sufficient number of classified images in the training set, the accuracy of the conventional CNN can reach 98.9% when classifying between melanoma cells and healthy skin cells, based only on their OPD maps. In contrast to classic computer vision algorithms, like support vector machine (SVM) or KNN, deep learning applied directly on the OPD maps automatically recognizes the important features that characterize cancer cells and healthy skin cells and uses them in order to distinguish between the different disease stages of the cell effectively. Hence, we eliminate the need in subjective and manual selection of features for optimization and enable rapid and efficient identification of individual cells without the need in staining. As a result, we obtain much better results, as shown in Tables 1 and 2. Indeed, SVM applied on a similar data have obtained 81% sensitivity and 83% specificity for discriminating between melanoma (Hs 895.T) and healthy skin (Hs 895.Sk) cells, and 82% sensitivity and 81% specificity for discriminating between primary (SW 480) and metastatic (SW 620) colorectal adenocarcinoma cells [21].

We also demonstrate the weakness of conventional CNN when the complexity of the data set classes increases or the available training set size decreases. We have shown that TOP-GAN, which is based on an integration of two learning approaches, transfer learning and GAN, is more robust for these cases. TOP-GAN is thus applied in the first time for the medical imaging field. By combining TOP-GAN with a familiar data augmentation method, we are able to increase the classification results more than other classification methods. Specifically, in the case of normal human skin cells (Hs 895.Sk) and a melanoma tumor cell from the skin (Hs 895.T) taken from the same individual, based on only 10 images for training, the classification results reach 98.96% sensitivity, 99.61% specificity and 99.054% accuracy. In the case of comparing between primary colorectal adenocarcinoma colon cells (SW 480) and metastatic from lymph node of colorectal adenocarcinoma cells (SW 620) taken from the same individual, based on 236 images for training, the classification results reach 93.01% sensitivity, 93.31% specificity and 90.6% accuracy. Thus, we eliminate the need to collect a large amount of classified data, which might be a great challenge in various medical and non-medical classification tasks. The GAN pre-training can last up to several hours, and requires a large number of unclassified images, in contrast to simpler methods like data augmentation. However, after pre-training the GAN on a general unclassified biological cell images (typically more accessible), the TOP-GAN training only depends on the size of the classified training set, which is relatively small, make the training process very quick. For example, for 160 images, the training process lasts 3.38 sec for TOP-GAN and 13 sec for MOBILE-NET, and for testing 40 images, it lasts 0.0226 sec for TOP-GAN and 0.62 sec for MOBILE-NET.

We believe that our new method will provide a valuable tool for an automatic classification process of individual cells for medical diagnosis in flow cytometry, as well as a new classification approach for medical images. In contrast to flow cytometry that obtains one accumulative measurement per cell, imaging flow cytometry allows measuring the cell morphology. Current imaging flow cytometers use cell staining, and allow imaging of up to several thousands of cells per second. In this paper, we presented proof-of-concept stain-free quantitative imaging for flow cytometry with much lower throughput of several cells per second. However, in principle, the techniques presented here can reach, with the suitable flow setups and fast cameras, to throughputs similar to the current imaging flow cytometers, but without cell staining and with the possibility of extracting more quantitative data from the cells measured, which create a better basis for machine learning classifiers.

References and Footnotes

[1] P. Paterlini-Brechot, NL. Benali, "Circulating tumor cells (CTC) detection: Clinical impact and future directions," *Cancer Lett*, vol. 253, no. 2, pp. 180–204, 2007.
[2] M. Yu, S. Stott, M. Toner, S. Maheswaran, DA. Haber, "Circulating tumor cells: Approaches to isolation and characterization," *J Cell Biol*, vol. 192, no. 3, pp. 373–382, 2011.
[3] E. Crowley, F. Di Nicolantonio, F. Loupakis, A. Bardelli, "Liquid biopsy: Monitoring cancer-genetics in the blood," *Nat Rev Clin Oncol*, vol. 10, no. 8, pp. 472–484, 2013.
[4] SF. Ibrahim, G. van den Engh, "Flow cytometry and cell sorting," *Adv Biochem Eng Biotechnol*, vol. 106, pp. 19–39, 2007.
[5] EC. Jensen, "Use of fluorescent probes: Their effect on cell biology and limitations," *Anat Rec*, vol. 295, no. 12, 2031–2036, 2012.
[6] PH. Wu, JM. Phillip, SB. Khatau, WC. Chen, J. Stirman, S. Rosseel, K. Tschudi, J. Van Patten, M. Wong, S. Gupta, "Evolution of cellular morpho-phenotypes in cancer metastasis," *Sci Rep*, vol. 5, 18437, 2015.
[7] D. Zink, AH. Fischer, JA. Nickerson, "Nuclear structure in cancer cell," *Nat Rev Cancer*, vol. 4, no. 9, pp. 677–687, 2004.
[8] P. Girshovitz, NT. Shaked, "Generalized cell morphological parameters based on interferometric phase microscopy and their application to cell life cycle characterization," *Biomed Opt Express*, vol. 3, no. 8, pp. 1757–1773, 2012.
[9] B. Rappaz, E. Cano, T. Colomb, J. Kühn, C. Depeursinge, V. Simanis, PJ. Magistretti, P. Marquet, "Noninvasive characterization of the fission yeast cell cycle by monitoring dry mass with digital holographic microscopy," *J Biomed Opt*, vol. 14, no. 3, 34049, 2009.
[10] NT. Shaked, "Quantitative phase microscopy of biological samples using a portable interferometer," *Opt Lett*, vol. 37, no. 11, pp. 2016–2019, 2012.
[11] P. Girshovitz, NT. Shaked, "Compact and portable low-coherence interferometer with off-axis geometry for quantitative phase microscopy and nanoscopy," *Opt Express*, vol. 21, no. 5, pp. 5701–5714, 2013.
[12] D. Roitshtain, NA. Turko, B. Javidi, NT. Shaked, "Flipping interferometry and its application for quantitative phase microscopy in a micro-channel," *Opt Lett*, vol. 41, no. 10, pp. 2354–2357, 2016.
[13] A. Nativ, NT. Shaked, "Compact interferometric module for full-field interferometric phase microscopy with low spatial coherence illumination," *Opt Lett*, vol. 42, no. 8, pp. 1492–1495, 2017.
[14] B. Javidi, I. Moon, S. K. Yeom, and E. Carapezza, "Three-dimensional imaging and recognition of microorganism using single-exposure on-line (SEOL) digital holography," *Optics Express*, vol. 13, no. 12, pp. 4492–4506, June 13, 2005.
[15] S. Yeom, I. Moon, B. Javidi, "Real-time 3D sensing, visualization and recognition of biological microorganisms," *Proceedings of the IEEE Journal*, vol. 94, no. 3, pp. 550-567, 2006.
[16] I. Moon, B. Javidi "3-D visualization and identification of biological microorganisms using partially temporal incoherent light in-line computational holographic imaging," *IEEE Transactions on Medical Imaging- (TMI)*, 27, 12, pp 1782-1790, 2008.
[17] E. Watanabe, T. Hoshiba, B. Javidi, B, "High-precision microscopic phase imaging without phase unwrapping for cancer cell identification," *Optics Letters*, vol. 38, pp. 1319–1321, 2013.
[18] C. Martinez-Torres, B. Laperrousaz, L. Berguiga, E. Boyer-Provera, J. Elezgaray, FE. Nicolini, V. Maguer-Satta, A. Arneodo, F. Argoul, "Deciphering the internal complexity of living cells with quantitative phase microscopy: A multiscale approach," *J Biomed*, vol. 20, no. 9, 96005, 2015.
[19] T. Go, JH. Kim, H. Byeon, SJ. Lee. (2018, April). "Machine learning-based in-line holographic sensing of unstained malaria-infected red blood cells," *Biophotonics*, vol. 11, e201800101, 2018.

[20] SK. Mirsky, I. Barnea, M. Levi, H. Greenspan, NT. Shaked, "Automated analysis of individual sperm cells using stain-free interferometric phase microscopy and machine learning," *Cytometry Part A,* vol. 91, no. 9, pp. 893-900, 2017.

[21] D. Roitshtain, L. Wolbromsky, E. Bal, H. Greenspan, LL. Satterwhite, NT. Shaked, "Quantitative phase microscopy spatial signatures of cancer cells," *Cytometry Part A,* vol. 91, no. 5, pp. 482–493, 2017.

[22] M. Hejna, A. Jorapur, JS. Song, RL. Judson, "High accuracy label-free classification of single-cell kinetic states from holographic cytometry of human melanoma cells," *Scientific Reports,* vol. 7, no. 1, 11943, 2017.

[23] J. Yoon, K. Min-hyeok, K. Kyoohyun, L. SangYun, K. Suk-Jo, P. YongKeun, "Identification of non-activated lymphocytes using three-dimensional refractive index tomography and machine learning," *Sci. Rep.*, vol. 7, no.1, 6654, 2017.

[24] H. S. Park, M. T. Rinehart, K. A. Walzer, J.-T. A. Chi, A. Wax, "Automated detection of P. falciparum using machine learning algorithms with quantitative phase images of unstained cells," *PLoS ONE*, vol. 11, no. 9, e0163045, 2016.

[25] VK. Lam, TC. Nguyen, BM. Chung, G. Nehmetallah, CB. Raub, "Quantitative assessment of cancer cell morphology and motility using telecentric digital holographic microscopy and machine learning," *Cytometry Part A,* vol. 93, no. 3, pp. 334–345, 2017.

[26] CL. Chen, A. Mahjoubfar, LC. Tai, IK. Blaby, A. Huang, KR. Niazi, B. Jalali, "Deep learning in label-free cell classification," *Sci Rep,* vol. 6, 21471, 2016.

[27] I. Moon, M. Daneshpanah, B. Javidi, A. Stern, "Automated three-dimensional identification and tracking of micro/nanobiological organisms by computational holographic microscopy," *Proc IEEE,* vol. 97, no. 6, pp. 990–1010, 2009.

[28] Y. Jo, S. Park, J. Jung, J. Yoon, H. Joo, M.H. Kim, S.J. Kang, M.C. Choi, S.Y. Lee, Y. Park, "Holographic deep learning for rapid optical screening of anthrax spores," *Sci Adv,* vol. 3, no. 8, e1700606, 2017.

[29] Y. Rivenson, T. Liu, Z. Wei, Y. Zhang, A. Ozcan, "PhaseStain: Digital staining of label-free quantitative phase microscopy images using deep learning," *arXiv preprint,* arXiv: 1807.07701, 2018.

[30] Y. Jo, H. Cho, S. Y. Lee, G. Choi, G. Kim, H.-s. Min, and Y. Park, "Quantitative phase imaging and artificial intelligence: A review," IEEE Journal of Selected Topics in Quantum Electronics, vol. 25, no. 1, pp. 1-14, 2019.

[31] Y. Rivenson, Y. Zhang, H. Gunaydin, D. Teng, and A. Ozcan, "Phase recovery and holographic image reconstruction using deep learning in neural networks," *arXiv preprint,* arXiv:1705.04286, 2017.

[32] O. Ronneberger, P. Fischer, T. Brox, "U-net: Convolutional networks for biomedical image segmentation," *Proc. Int. Conf. Medical Image Comput. Comput.-Assisted Intervention*, pp. 234–241, 2015.

[33] Y. LeCun, Y. Bengio, G. Hinton, "Deep learning," *Nature,* vol. 521, pp. 436–444, 2015.

[34] M. Oquab, L. Bottou, I. Laptev, J. Sivic, "Learning and transferring mid-level image representations using convolutional neural networks," in *Proc. IEEE Conf. CVPR*, 2014, pp. 1717–1724.

[35] D. Erhan, P.A. Manzagol, Y. Bengio, S. Bengio, P. Vincent, "The difficulty of training deep architectures and the effect of unsupervised pre-training, " " in *Proc. ICAIS*, 2009, pp. 153–160.

[36] N. Tajbakhsh, J.Y. Shin, S.R. Gurudu, R.T. Hurst, C.B. Kendall, M.B. Gotway, J. Liang, "Convolutional neural networks for medical image analysis: Full training or fine tuning," *IEEE Tran Med Imaging*, vol. 35, no. 5, pp. 1299–1312, 2016.

[37] J. Shi, S. Zhou, X. Liu, Q. Zhang, M. Lu, T. Wang, "Stacked deep polynomial network based representation learning for tumor classification with small ultrasound image dataset," *Neurocomputing*, vol. 194, pp. 87–94, 2016.

[38] G. Litjens, "A survey on deep learning in medical image analysis," *Med Image Anal*, vol. 42, pp. 60–88, 2017.

[39] A. Krizhevsky, I. Sutskever, and G. E. Hinton, "Imagenet classification with deep convolutional neural networks," *NIPS*, pp. 1097–1105, 2012.

[40] H.C. Shin, H.R. Roth H. R, M. Gao, L. Lu, Z. Xu, I. Nogues I, J. Yao, D. Mollura, R.M. Summers, "Deep convolutional neural networks for computer-aided detection: CNN architectures, dataset characteristics and transfer learning," *IEEE Tran Med Imaging*, vol. 35, no. 5, pp. 1285–1298, 2016.

[41] Y. Bar, I. Diamant, L. Wolf, H. Greenspan, "Deep learning with non-medical training used for chest pathology identification," in *Proc. SPIE Med Imag,* 2015, pp. 94140V–94140V.

[42] A. Howard, M. Zhu, B. Chen, D. Kalenichenko, W. Wang, T. Weyand, M. Andreetto, and H. Adam. "MobileNets: Efficient convolutional neural networks for mobile vision applications," *arXiv preprint,* arXiv:1704.04861, 2017.

[43] I. Goodfellow, J. Pouget-Abadie, M. Mirza, B. Xu, D. Warde-Farley, S. Ozair, A. Courville , Y. Bengio, "Generative adversarial nets," *NIPS*, pp. 2672–2680, 2014.

[44] A. Radford, L. Metz, S. Chintala, "Unsupervised representation learning with deep convolutional generative adversarial networks," *arXiv:* 1511.06434, 2015.

[45] M. Frid-Adar, I. Diamant, E. Klang, M. Amitai, J. Goldberger, H. Greenspan, "GAN-based synthetic medical image augmentation for increased CNN performance in liver lesion classification," *arXiv preprint,* arXiv: 1803.01229, 2018.

[46] A. Odena, "Semi-supervised learning with generative adversarial networks," *arXiv preprint,* arXiv:1606.01583, 2016.

[47] T. Salimans, I. Goodfellow, W. Zaremba, V. Cheung, A. Radford, X. Chen, "Improved techniques for training gans," *arXiv preprint,* arXiv:1606.03498, 2016.

[48] C. Vondrick, H. Pirsiavash, A. Torralba, "Generating videos with scene dynamics," *NIPS*, pp. 613–621, 2016.

[49] U. Ahsan, C. Sun, I. Essa, "Discrimnet: Semi-supervised action recognition from videos using generative adversarial networks," *arXiv preprint,* arXiv: 1801.07230, 2018.

[50] I. Moon, "Identification of Malaria Infected Red Blood Cells via Digital Shearing Interferometry and Statistical Inference," *IEEE Photonics Journal*, vol. 5, no. 5, 6900207, 2013.

[51] P. Girshovitz, NT. Shaked, "Fast phase processing in off-axis holography using multiplexing with complex encoding and live-cell fluctuation map calculation in real-time," *Opt Express*, vol. 23, no.7, pp. 8773–8787, 2015.

[52] P. Teare, M. Fishman, O. Benzaquen, E. Toledano, E. Elnekave, "Malignancy detection on mammography using dual deep convolutional neural networks and genetically discovered false color input enhancement," *J. Digit. Imaging*, vol. 30, no. 4, pp. 499–505, 2017.

[53] O. Boiman, E. Shechtman, and M. Irani, "In defense of nearest-neighbor based image classification," in *CVPR08*, 2008, pp. 1–8.

[54] G. Cohen, G. Sapiro, R. Giryes, "DNN or K-NN that is the Generalize vs. Memorize Question," *arXiv preprint,* arXiv: 1805.06822, 2018.

[55] M. Abadi, A. Agarwal, P. Barham, E. Brevdo, Z. Chen, C. Citro, G. S. Corrado, A. Davis, J. Dean, M. Devin, "Tensorflow: Large-scale machine learning on heterogeneous distributed systems," *arXiv preprint,* arXiv:1603.04467, 2016.

[56] N. Srivastava, G. E. Hinton, A. Krizhevsky, I. Sutskever, R. Salakhutdinov, "Dropout: a simple way to prevent neural networks from overfitting," *Journal of Machine Learning Research*, vol. 15, no. 1, pp. 1929–1958, 2014.